\documentclass[journal]{IEEEtran}

\usepackage[pdftex]{graphicx}
\usepackage{amsmath}
\usepackage{amssymb}
\usepackage{chngcntr}
\usepackage{float}
\usepackage{ifpdf}
\usepackage{subfigure}
\usepackage{xcolor}

\usepackage{cuted}

\usepackage[normalem]{ulem}

\usepackage{cite}
\usepackage{amsmath,amssymb}

\usepackage{dsfont}

\def\half{{\textstyle{1\over 2}}}

\def\Eqref#1{Eq.~\eqref{#1}}

\def\be#1\ee{\begin{equation}#1\end{equation}}
\def\bea#1\eea{\begin{align}#1\end{align}}
\def\bse#1\ese{\begin{subequations}#1\end{subequations}}

\def\1#1{{\hat{{\boldsymbol{#1}}}}}                                 		%hat: bold-symbol
\def\2#1{\hat{#1}}                                              		   		%widehat
\def\3#1{{\mathbf{#1}}}                                             	   		%bold-roman
\def\4#1{{\boldsymbol{#1}}}                                            		%bold-symbol
\def\5#1{{\mathcal#1}}                                                            		%caligraphy
\def\6#1{\bar{#1}}                                                           		%over-bar
\def\7#1{{{#1}}}                                    		%small
\def\8#1{\widetilde{#1}}
\def\9#1{\check{#1}}
\def\+#1{{\overset{{\scriptscriptstyle +}}{#1}{}}}                  		%superplus
\def\b+#1{{\overset{{\scriptscriptstyle +}}{\mathbf{#1}}{}}}        	%superplus  bold-roman
\def\g+#1{{\overset{{\scriptscriptstyle +}}{\boldsymbol{#1}}{}}}    %superplus bold-symbol
\definecolor{dark-green}{rgb}{0.278,0.7,0.4}                    %Color = definition of dark green

\definecolor{my-brown}{rgb}{0.69,0.247,0.13}                    %Color = definition of my brown

\definecolor{my-purple}{rgb}{0.47,0.12,0.46}                    %Color = definition of my purple

\definecolor{my-greenblue}{rgb}{0.129,0.313,0.419}              %Color = definition of my green-blue

\definecolor{my-orange}{rgb}{1,0.5,0.25}                        %Color = definition of my orange

\definecolor{my-red}{rgb}{0.745,0,0.2117}                       %Color = definition of my red

\definecolor{my-gray}{rgb}{0.5,0.5,0.5}                         %Color = definition of my gray

\definecolor{my-dark-blue}{rgb}{0.1,0.1,0.7}                    %Color = definition of my dark blue

\definecolor{my-indigo}{rgb}{0.29,0.0,0.51}                    %Color = definition of my indigo

\definecolor{my-blue-violet}{rgb}{0.54, 0.17, 0.89}

\definecolor{byzantine}{rgb}{0.74, 0.2, 0.64}
\begin{document}
%
% paper title
% Titles are generally capitalized except for words such as a, an, and, as,
% at, but, by, for, in, nor, of, on, or, the, to and up, which are usually
% not capitalized unless they are the first or last word of the title.
% Linebreaks \\ can be used within to get better formatting as desired.
% Do not put math or special symbols in the title.
%\title{Indirect Time-Modulation of Antennas: a Venue Beyond  Size dependent bounds}
\title{Parasitic Element Time-Modulation for Enhanced Effective Inter-Antenna Coupling: Utilization for Improved Gain-Bandwidth}
%

%Parasitic Element Time-Modulation for Enhanced Effective Inter-Antenna Coupling: Utilization for Improved Gain-Bandwidth

%Enhanced Effective Inter-Antenna Coupling by Time-Modulation of a Parasitic Element: Utilization for Improved Gain-Bandwidth

% Enhanced Effective Inter-Antenna Coupling through Indirect Time-Modulation: Utilization for Improved Gain-Bandwidth

%
% author names and IEEE memberships
% note positions of commas and nonbreaking spaces ( ~ ) LaTeX will not break
% a structure at a ~ so this keeps an author's name from being broken across
% two lines.
% use \thanks{} to gain access to the first footnote area
% a separate \thanks must be used for each paragraph as LaTeX2e's \thanks
% was not built to handle multiple paragraphs
%

\author{Amir~Shlivinski,~\IEEEmembership{Senior Member,~IEEE,}, Yakir~Hadad,~\IEEEmembership{Senior Member,~IEEE,}
% <-this % stops a space
\thanks{A. Shlivinski is with the School of Electrical and Computer Engineering, Ben Gurion University of the Negev, Beer Sheva, Israel, 84105, e-mail: amirshli@ee.tau.ac.il.}
\thanks{Y. Hadad is with the School
of Electrical Engineering, Tel Aviv University, Tel Aviv, 69978 Israel, e-mail: hadady@eng.tau.ac.il}% <-this % stops a space
\thanks{This work was supported by the Israel Science Foundation
under Grant 1353/19}}
\maketitle

% As a general rule, do not put math, special symbols or citations
% in the abstract or keywords.
\begin{abstract}
Time variation has been recently introduced as an additional degree of freedom for wave engineering, that enables going beyond the performances that are expected by linear time-invariant (LTI) systems. In this paper, we introduce the concept of indirect time-modulation of antennas using an add-on time-varying scatterer (parasitic element) that gives rise to an inherent feedback mechanism via the airborne wave system. As opposed to a direct modulated system where a time-dependent element is in contact with the other elements,  in an indirect time modulation scheme \emph{no} direct physical contact between the original LTI network and the time-varying add-on scatterer is needed, thus leading to additional flexibility in the design. Using indirect time modulation we demonstrate enhanced effective coupling between remote antenna elements, and the possibility to outperform the gain-bandwidth achieved for the same antenna structure but without time-modulation.
\end{abstract}

% Note that keywords are not normally used for peerreview papers.

% For peer review papers, you can put extra information on the cover
% page as needed:
% \ifCLASSOPTIONpeerreview
% \begin{center} \bfseries EDICS Category: 3-BBND \end{center}
% \fi
%
% For peerreview papers, this IEEEtran command inserts a page break and
% creates the second title. It will be ignored for other modes.
\IEEEpeerreviewmaketitle

%%%%%%%%%%%%%%%%%%%%%%%%%%
\section{Introduction}

The coupling between adjacent small antennas can be described in terms of the mutual impedance between the antennas. This, typically, decays asymptotically with the distance between the antennas like $\sim (l/\lambda)^2/(d/\lambda)$ where $l$ denotes the dipoles size and $d$ the spacing between the dipoles. This means that short dipole antennas with $l\sim 0.1\lambda$ separated by a distance $d>2-3\lambda$ can practically be considered as isolated.
While often, one seeks to isolate between the antenna elements, on certain occasions, the coupling between the antenna elements or the coupling between the antenna and its close vicinity, can be utilized in order to tailor particular radiative and feeding properties such as high directivity and super-gain \cite{Schelkunoff1943,Ziolkowski2006,Yaghjian2008,Shi2022}.
As opposed to the antenna directivity which is in principle an unbounded quantity (although, practical implementations of super-gain are obviously challenging), the antenna quality factor (Q), and the gain-bandwidth product are fundamentally bounded by Chu \cite{Chu1948}.
These bounds are intrinsic to the antenna as a radiation device without regarding any matching with the feeding guiding structure. A maximization of the reflection coefficient bandwidth at a feed terminal is achievable by proper additional matching network that may be regarded as shielded from radiation as suggested by the existence of the Bode-Fano  bound for such LTI systems \cite{Bode,Fano}. This observation sets a clear distinction between Chu's and the Bode-Fano bounds. The concept of the Chu bound was later modified to include the interface properties (matching) with the feeding network at the TL terminal  \cite{Gustafsson2016}.
It is well known that improved performance can be obtained using non-Foster designs \cite{Sussman2009, Chen2013, Soric2014}, with penalty in terms of increased noise figure \cite{Pozar} (see \cite{Shi2019} for an improved design) and potential instabilities \cite{Stearns2011,Stearns2012}.
In this context, ref.~\cite{Yaghjian2018} shows that with highly dispersive and lossy (non-hermitian) materials the Chu bound on the minimal antenna's Q is invalid due to the improper definition of the Q-energy in highly dispersive materials versus in a quasi-static circuit model with dispersion-free equivalent elements.

In this work we suggest a possible venue to enhance the antenna dispersion and its non-hermiticity by introducing simultaneously additional resonances in the antenna system with parametric time-modulation.
Specifically, here we explore the effect of indirect time-modulation on the antenna characteristics. The term \emph{indirect} refers to the use of an additional adjacent time modulated scatterer (parasitic element) that has \emph{no physical contact} with the other elements. This parasitic element is used in order to change the performance of the system, see in Fig.~\ref{Fig1}. Indirect modulation, stands as opposed to direct time-modulation schemes which have been used extensively recently for various applications, from impedance matching \cite{Shlivinski2018,Hadad2020}, through manipulation on bound states in the continuum \cite{Hayran2021}, absorption \cite{Sievenpiper2020, Li2021, Firestein2022, Yang2022, Monticone2023}, and radiation \cite{Galejs1963, Wang2004_1,Wang2004_2, Wang2014, Hadad2016, Li2019, Mekawy2021,Mostafa2023}.
The indirect time-modulation scheme is specifically implemented by introducing a nearby add-on time-modulated scattering element.
By doing so,  the adjacent added time-modulated scattering element simultaneously modifies  \emph{both} the \emph{matching properties at the feed terminal} of the antenna and its \emph{radiation properties at the radiation terminal}, in contrast with direct time-modulation that can affect only the matching properties at the feed terminal.
Interestingly, by using the parametric modulation with a properly optimized antenna array, we show that the effective mutual coupling between the elements increases significantly, leading to dominant mutual effects at large distances. Therefore, enabling the design of dispersive non-hermitian response.
By following this approach, as an example,  we analyze the \emph{gain-bandwidth product} of a compact three short dipole antenna array system with indirect coupled time modulated scattering element, and show  that the system achieves superior performance by the introduction of indirect time modulation compared to what is achievable for the same antenna structure but without time modulation.

In the following discussion we are primarily interested with the effect of LTV vs LTI indirect coupling on the performances of an antenna system. Hence below we first describe the concept of indirect time-modulation, followed by numerical demonstrations that provide a comparison between LTI and LTV cases. The detailed  mathematical derivation that is required to model the antenna system is deferred to the appendix.

\section{Indirect versus direct coupling}
%%%%%%%%%%%%%%%%%%%%%%%%%%
%{\color{red} %%Amir: I dont understand this sentence
%Following the discussion in the introduction specifically for the case of a radiating system where both matching to a feedline in conjunction to maximal radiation are of interest.}

Assuming that the radiating element is stationary in time, the application  of time  modulation within the system can be carried via two different layouts. The first, is by a ``direct coupling' scheme where a linear time-variant circuit element is embedded in a  matching network that connects the antenna and the feeding transmission line, see in Fig.~\ref{Fig1}(a). This was suggested in the past, e.g., in \cite{Galejs1963} and more recently in \cite{Li2019,Mekawy2021,Mostafa2023}. The second layout is that of an `indirect coupling' where an additional secondary (auxiliary) source-less linear time-variant (LTV) scattering system is placed in a proximity to the primary antenna such that the two systems are electromagnetically coupled and \emph{the radiating properties of the combined system} as well as the \emph{terminal matching} are modified, see in Fig.~\ref{Fig1}(b). The two schemes are clearly different in their operation. While in the `direct coupling' the LTV matching network provides matching of the radiating element to the feeding transmission-line (TL) terminal, the radiation characteristics remain the same  (radiation pattern, impedance, etc.) as of those of LTI system. Opposingly, in the `indirect coupling' setup both the matching to the feeding TL and the \emph{radiation} properties of the antenna system are altered by the close proximity of the two radiating systems. As a consequence, a striking difference between the layouts is that in the indirect coupling case the two adjacent radiating systems in conjunction with time variation are interacting as a \emph{feedback loop} that, with a proper balance between the electromagnetic coupling and the time variation, can provide both wideband matching and antenna gain increase while maintaining  the stability of the network. The following discussion seeks to demonstrate that by adding some time-variation that is indirectly coupled with a radiating element,  better performance can be obtained when compared to the same system however stationary in time.

%%%%%%%%%%%%%%%%%%%%%%%%
\section{A 3-dipole Yagi-Uda-like array}
%%%%%%%%%%%%%%%%%%%%%%%%
{To demonstrate this concept we consider the transmission of a finite bandwidth real signal with frequency spectrum shown in Fig.~\ref{Fig1}(c) centered around $\omega_0$, a central radian frequency.   We are mainly focused on exploring the abstract measure of the \emph{gain-bandwidth} product for the indirect time-modulation network, see in Fig.~\ref{Fig1}(b), and specifically for the case of the three antennas array in Fig.~\ref{Fig1}(d), consisting of one primary element with LTI inductor matching, $L_1$, that is connected to a feeding TL, and two secondaries (parasitic) elements in a Yagi-Uda like antenna style. Regarding the two parasitic elements, \#3 is loaded by an LTI inductor $L_3$, and \#2 is loaded with a possible periodically time-modulated inductor, $L_2$.
%%%%%%%%
%\be
%L_2(t)=L_{20} \bigl (1+2\lambda_1 \cos(\omega_c t)  = L_{20}\sum_{\ell=-1}^{1} \lambda_\ell e^{\ell\omega_c t},
%\label{L2_t}
%\ee
%%%%%%%%
%with $\lambda_1=\lambda_{-1} <0.5$, $\lambda_0=1$, and ${\rm Im}\{\lambda_{\pm1}\} =0$ and where $\omega_m=2\omega_r$ is the modulation frequency with $\omega_r$ a reference frequency.
%{\color{red} It is interesting to note that this small array resembles a Yagi-Uda like antenna}}

%%%%%%%%%%%%%%%%%%%%%%%%%%%
\begin{figure}[ptb]
\centering
\vspace{-0.0cm} \includegraphics[width=8cm]{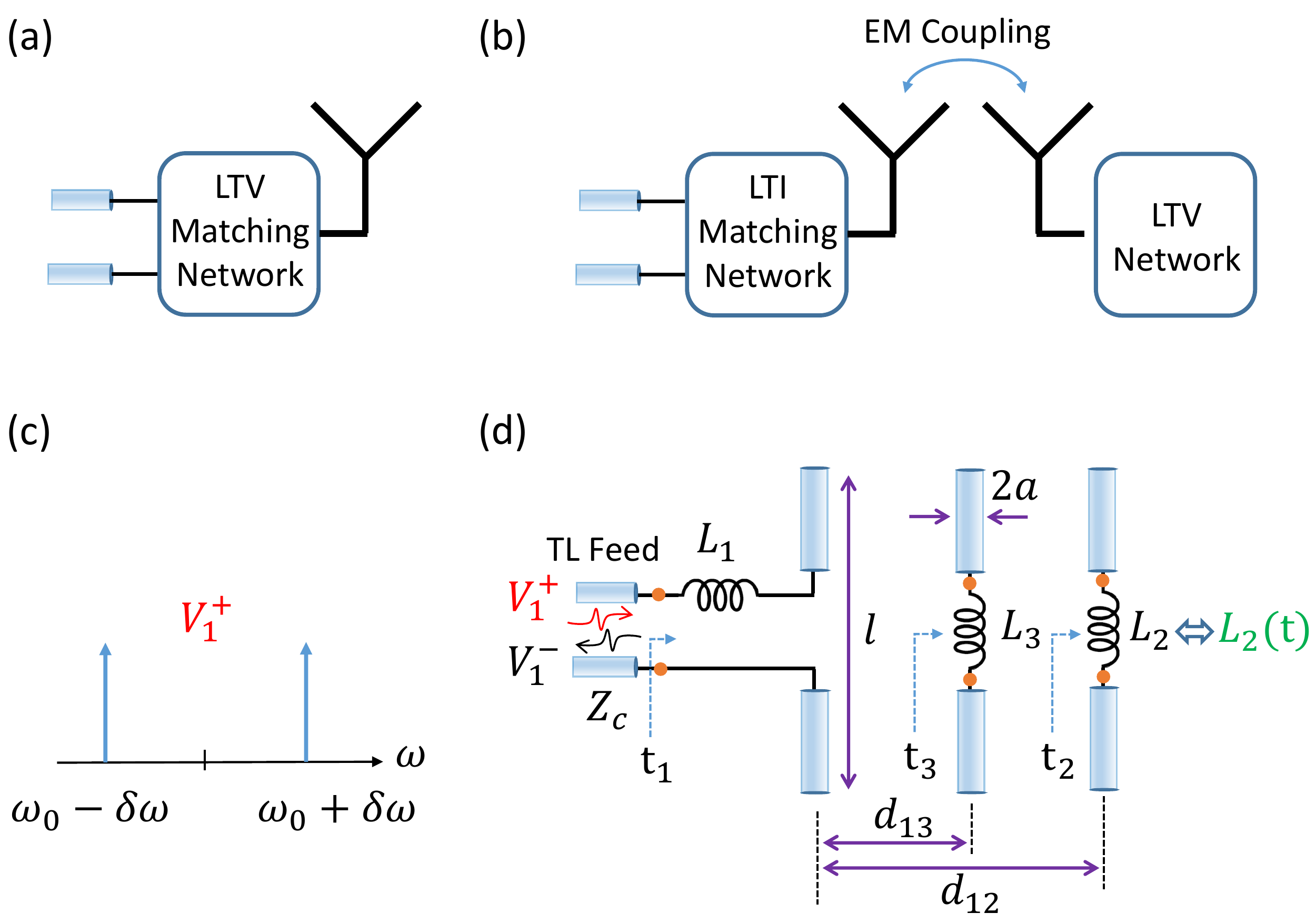} \vspace{-0.0cm}%
\caption{ A scheme of (a) direct time-modulation versus (b) indirect time modulation. (c) The exciting signal, $V_1^+$, spectrum (positive frequencies are shown). The signal bandwidth $2\delta\omega$ around $\omega_0$. (d) A 3-dipole antenna array. The inductance $L_1$ is fixed to place dipole \#1 at resonance in the absence of the parasitic elements, the dipoles share a fixed identical length $l$ and diameter $2a$. The array size $d_{12}$ is fixed as well. The remaining parameters $L_2,L_3$ and $d_{13}$ are used for the optimization. The 3-dipole system is analyzed as a three-port network with port terminals $t_1$, $t_2$, and $t_3$. The terminal points are marked with orange circles. The $t_1$ terminal points are connected to the feed TL, the $t_2$ and $t_3$ terminal points are connected to the inductors $L_2$ and $L_3$, respectively.}
\label{Fig1}%
\end{figure}
%%%%%%%%%%%%%%%%%%%%%%%%%%%

%The three dipole elements are electrically short with length of $l \ll \lambda_0$, where  $\lambda_0=2\pi c/\omega_0$ denotes the wavelength at some central radian frequency $\omega_0$, and with $c$ denoting the speed of light in vacuum. Additional array parameters include the loading inductances $L_1$, $L_2,$ and $L_3$, the total array size $d_{12}$ which is the distance between the primary dipole \#1 and the time-modulated parasitic dipole \#2, as well as the distance between dipoles \#1 and \#3, $d_{13}$.  The inductance $L_1$ is tuned such that dipole \#1 is at resonance in the absence of the parasitic elements (\#2, and \#3) at $\omega_0$.

%Consider first the array in Fig.~\ref{Fig1}(d) in the absence of any time modulation in element \#2 {\color{red} (a pure LTI system)}. For a fixed array size $d_{12}$, and dipole length $l$,  one may optimize the antenna array performance over the remaining three parameters $L_2$, $L_3$, and $d_{13}$ to achieve certain goals.
%
\subsection{Generalized definition for the gain-bandwidth}
Given an antenna system whether LTI or LTV with direct or indirect coupling, there are many possible performance measures that the system can be optimize for. Of particular interest are ($i$) maximizing the \emph{reflection coefficient bandwidth}, that is, the bandwidth as seen at the input port, ($ii$) maximizing the \emph{gain} of the array, or ($iii$) maximizing the \emph{gain-bandwidth product} of the antenna. While \emph{reflection coefficient bandwidth} and an antenna \emph{gain} are relatively easy to define for a system, the \emph{gain-bandwidth product} is more intricate to define since it involves  properties at two different type of terminals (lumped and dustributed) and for a system that consists of multiple coupled resonant elements and thus its frequency response is more complex than a second order Lorentzian. Moreover it becomes even more subtle to define in the case of LTV system where power can flow into/out of the system also from the ``modulating element port'' and at different frequency harmonics beyond the fundamental harmonic at $\omega_0$. To that end and since the gain-bandwidth product is of our main interest here we start by providing its definition as will further be used.
%
%Before moving on, we should pay some attention to the way we define the gain bandwidth for the antenna array in Fig.~\ref{Fig1}(d). It is important to note that the Yagi-Uda like antenna  consists of multiple coupled resonant elements and thus its frequency response is more complex than a second order Lorentzian type. This, and the fact that it is time-modulated, introduce an issue with the use of the conventional gain-bandwidth product.
%
To overcome the difficulties listed above, we introduce a quality measure that closely resembles, in its essence, and generalizes the \emph{gain-bandwidth product} associated with second order narrowband systems, but fits into a wideband signal framework and that can be applied also with LTV systems,
%%%%%%%%%%%%%%
\begin{equation}\label{GainBW}
GBW=\int\limits_{0}^{``\infty"} \left (1-\left |\Gamma_0\right |^2 \right ) \left [ 4\pi {\max_{\Omega} U_0(\Omega)}/{P_{in}}\right] d\delta_\omega,
\end{equation}
%%%%%%%%%%%%%%
%
where $U_0$ is the radiation intensity due to the \emph{fundamental harmonic contribution} around the central radian frequency $\omega_0$ and at the spatial direction $\Omega=(\sin \theta \cos\phi, \sin \theta \sin\phi, \cos \theta)$ ($0 \le \theta \le \pi$, $0 \le \phi < 2\pi$), $\Gamma_0$ is the reflection coefficient at the source terminal (at the TL source feed terminal in Fig.~\ref{Fig1}(d)) due to the \emph{fundamental harmonic}, $P_{in}=P_1+P_2$ is the power \emph{strictly  entering} the antenna terminals. The contribution $P_1$ is the net power (over \emph{all harmonic contributions}) flowing into the antenna system from the source terminal, $P_2$ is the actual power, over \emph{all harmonic contributions}, \emph{entering} antenna \#2 terminal ($P_2>0$) due to the inductance modulation when applied.
Note that for the calculation of the radiation intensity in Eq.~(\ref{GainBW}) we assumed a sinusoidal current on each of the single short antennas \cite{KrausAntennas,Balanis}. Once the antennas terminal currents are calculated (see Appendixes \ref{SingleTone} and \ref{Twotone}) the radiated field of the three-antennas array can be easily calculated and henceforth the radiation intensity.
Note that in Eq.~(\ref{GainBW}), the upper limit of the integration, $``\infty"$, should be understood such that it covers the frequency range where the system modeling still applies.

In view of this definition in Eq.~(\ref{GainBW}), it should be noted: $(i)$ The term ``bandwidth'' is related to the frequency bandwidth of the reflection coefficient at the actual system's ``source'' input terminal while the gain is defined with respect to all power sources in the system, namely, the actual source via the input terminal (terminal \#1) and additional ``parametric'' sources due to the possible modulation of the inductance $L_2$ (at terminal \#2); $(ii)$ since the LTV system is periodically time modulated, the signals in the system, i.e., the voltages and currents at the antennas' terminals, the induced currents on the antennas, and the radiated fields, are composed of many temporal harmonics at frequencies that are mixture of $\omega_m$, the modulation radian frequency,  and $\omega_0 \pm \delta_\omega$ that contribute to $P_2$; $(iii)$ with a strict second order time-harmonic LTI system, the definition in Eq.~(\ref{GainBW}) approaches a standard gain-bandwidth product figure of merit, to see that consider, for example, the special case where an antenna is fully matched within a certain bandwidth $BW_0$ (i.e., zero reflection coefficient) and has a gain $G_0$ to give that the gain-bandwidth is indeed $\mbox{GBW}=G_0 \times BW_0$.
$(iv)$ Furthermore, in light of items $(i)$ and $(ii)$, it should be stressed that the definition in Eq.~(\ref{GainBW}) in the way we apply it is most conservative since for the radiating power in the numerator we take only the power at the fundamental harmonic - at which we wish to transmit the signal, while for the incoming power in the denominator we take all time harmonics that are practically excited at the system.
%
%$(iv)$ Lastly we stress that the gain-bandwidth definition in Eq.~(\ref{GainBW}) is used to derive all the results shown in Fig.~\ref{Fig2}(a) for the stationary and time modulated cases.

%{\color{teal} To demonstrate this concept we consider the transmission of a finite bandwidth real signal with frequency spectrum shown in Fig.~\ref{Fig1}(c).   We are mainly focused on exploring the abstract measure of the \emph{gain-bandwidth} product for the indirect time-modulation network that is shown in Fig.~\ref{Fig1}(d),  a  three antennas array, consisting of one primary element with LTI matching that is connected to a feeding TL, and two secondaries (parasitic) elements, \#3 is a LTI element, and \#2 is loaded  with a time-modulated inductor. {\color{red} This small array resembles a Yagi-Uda like antenna}}

\subsection{Gain-bandwidth optimization: LTI vs LTV}

We explore and demonstrate the properties of gain-bandwidth product with the compact array in Fig.~\ref{Fig1}(d). The three dipole elements are electrically short with length of $ \ll  2\pi c/\omega_0$,  where $2\pi c/\omega_0$ denotes the wavelength the central radian frequency $\omega_0$, and with $c$ denoting the speed of light in vacuum. Additional array parameters include the loading inductances $L_1$, $L_2,$ and $L_3$, the total array size $d_{12}$ which is the distance between the primary dipole \#1 and the time-modulated parasitic dipole \#2, as well as the distance between dipoles \#1 and \#3, $d_{13}$.  The LTI inductance $L_1$ is tuned such that dipole \#1 is at resonance in the absence of the parasitic elements (\#2, and \#3) at $\omega_0$.The inductances $L_3$ is LTI and $L_2$ is generally a LTV inductance
%%%%%%%
\be
L_2(t)=L_{0} \bigl (1+m \cos(\omega_m t) \bigl), \qquad 0 \le m <1
\label{L2_ta}
\ee
%%%%%%%
where $L_0$ is the nominal inductance value and $\omega_m$ is the modulation frequency.
For formulation of the array behaviour, that accounts for the mutual coupling between the array elements, including multiple interactions between the terminals, please refer to Appendix \ref{Prelimenaries} below.

%
%In formulating and analyzing the array behavior, see the detailed discussion and derivation in the appendix section, we assumed a simplified model where the impedance matrix of the three elements array was constructed using the impedance matrixes of interactions of pairs of antennas ( $\#1-\#2$, $\#1-\#3$ and $\#2-\#3$) and considering also multiple interactions between the terminals.}

Consider, first, the array in Fig.~\ref{Fig1}(d) in the absence of any time modulation in element \#2 (a pure LTI system with $m=0$ in \eqref{L2_ta}, namely, $L_2=L_0$). For a fixed array size $d_{12}$, and dipole length $l$,  one may optimize the antenna array performance over the remaining three parameters $L_2$, $L_3$, and $d_{13}$ to achieve for certain goals as discussed above. Here we perform this optimization to maximize the gain-bandwidth product. %Theses three optimizations were discussed in Chu's paper for a general LTI antenna that is enclosed within a sphere of radius $a$ to give a set of general performance bounds.
Such optimization yields the best performances for that 3 antennas system and hence maybe considered as tight LTI bounds under the assumption that all the elements in the array system are linear and time-invariant.
As an example for such a tight LTI bound, the maximally attainable gain-bandwidth, GBW of Eq.~\eqref{GainBW} for the three antennas array in Fig.~\ref{Fig1}(d) is numerically calculated and shown in Fig.~\ref{Fig2} by the dashed blue line as a function of the array size $d_{12}$. Notably, as the array size exceeds $\sim5\lambda_0$ the gain-bandwidth reaches a plateau value that suggests that the effect of the coupling between the elements becomes negligible and the GBW is that of a single ``isolated'' element.
%
%equals to the gain-bandwidth, Eq.~\eqref{GainBW}, of the primary element \#1 when isolated in space. This is an expected result since above this array size the mutual coupling between the elements becomes negligible.

%
%It is important to note that since this antenna system is composed of multiple resonant elements its frequency response is more complicated than a mere second order Lorentzian type. Therefore, the gain-bandwidth product should be generalized as in Eq.~(\ref{GainBW}) to be applied in this case for wideband signals, and later also to time-dependent systems.

What should be expected if we take the same class of antenna systems, but enable time modulation of one of its elements? To explore this question, we replace the stationary inductance $L_2=L_0$ with a time modulated inductance $L_2(t)$, as shown in Fig.~\ref{Fig1}(d), with $0 \le m<1$ as in \eqref{L2_ta}  $\omega_m= 2\omega_0$ (i.e. the modulation frequency which is set to be twice of the center frequency of the modulated input signal, see Fig.~\ref{Fig1}(c)). Thus, the time modulation is essentially a \emph{parametric modulation that may significantly pump energy into or sink energy off the radiating system}. For a detailed analysis of the antenna array under time-modulation see in the appendix.
In order to  derive the maximally attainable gain-bandwidth of the time-modulated antenna array we optimize over the four parameters, $L_0$, $m$, that represent the time-modulated inductance $L_2(t)$, and over $d_{13}$, and $L_3$ for a given $d_{12}$.
\begin{figure}[]
\centering
\includegraphics[width=8cm]{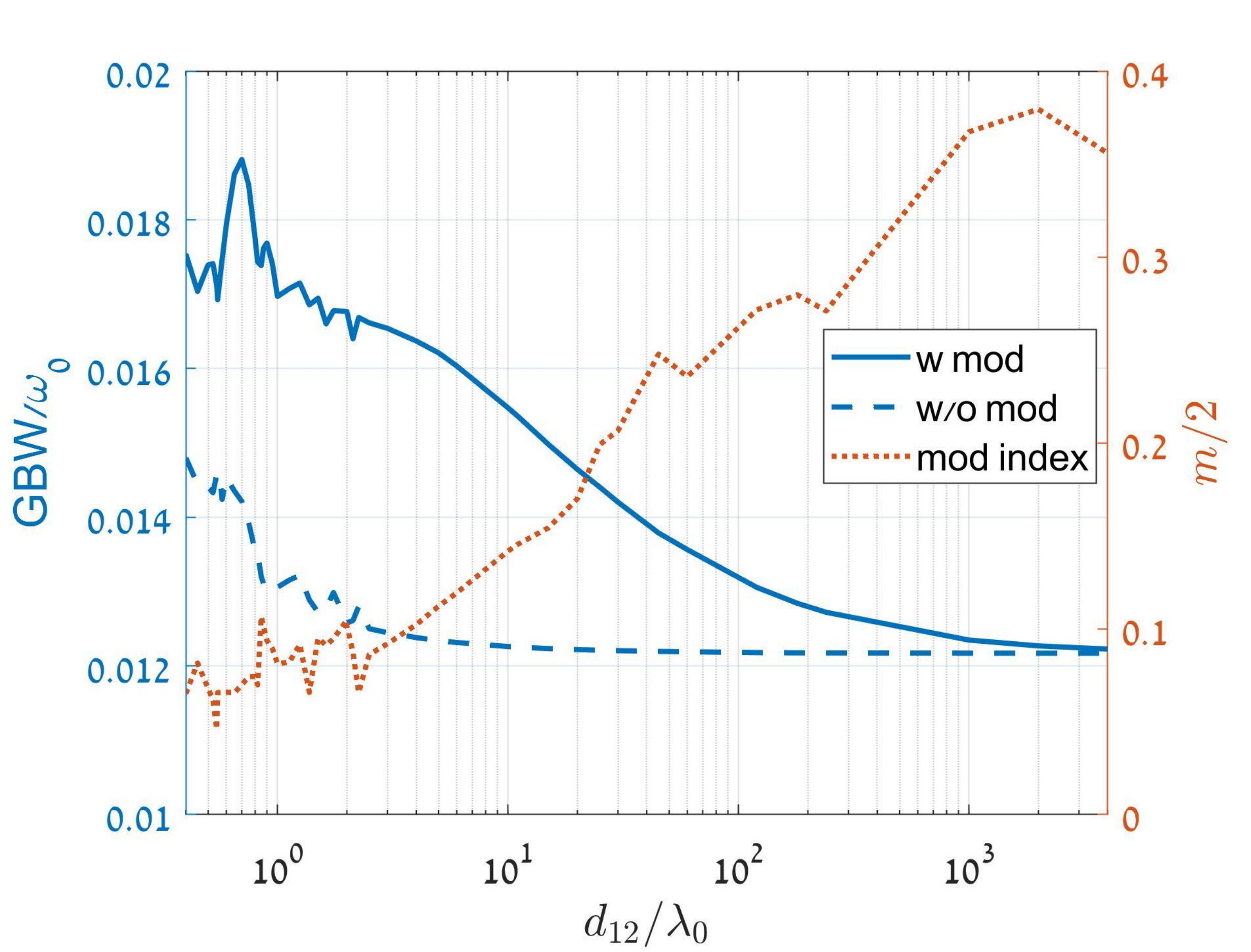}
\caption{The optimal $\mbox{GBW}/\omega_0$ for both the modulated and stationary systems (blue solid and dashed lines, respectively) and the modulation index $m/2$ (red line) as a function of the array size $d_{12}$}
\label{Fig2}
\end{figure}

%%%%%%%%%%%%%%%%%%%
%

The optimized GBW for the time-modulated case is shown in Fig.~\ref{Fig2} by the blue lines; dashed - for the LTI case, and cont. - for the LTV case. The associated modulation index $m$ that is required to get the optimized performance is shown in red. It can be readily noted that the time-modulated system (LTV) gives larger values of the GBW than those obtained with the stationary system (LTI), for any array size $d_{12}$, going from small, $d_{12}\sim0.4\lambda$, to an extremely large, $d_{12}\sim10^3\lambda_0$. This wide range of array sizes was set in order to explore the effect of modulation as a feedback mechanism that compensates over the weak electromagnetic coupling that is imposed by the large distance between the radiating elements.
We note that for $d_{12}/\lambda_0 \gtrsim 2$,  the GBW is a monotonically decreasing function of $d_{12}$ whereas for $d_{12} \lesssim 2$, the behavior is more ``undulating'' with local and global peaks at different $d_{12}$ values. This behavior, that is observed in both the modulated and stationary systems, is attributed to the relatively strong inter-antennas coupling effect between the antennas when in close proximity that is further increased by the time modulation. The strong coupling effect is reduced as antenna \#2 becomes more distant, for which the modulation index increases as an inherent compensation mechanism. Nevertheless, as seen in Fig.~\ref{Fig2} while the electromagnetic coupling becomes practically negligible for array sizes larger than $\sim5\lambda_0$ and therefore the gain-bandwidth reduces to that of an isolated primary antenna, even in the presence of time modulation, the effect of the coupling is clear for much larger array sizes, as seen by the fact that the GBW in the modulated case is larger (and decays somewhat slower with the increase in $d_{12}$) than in the non-modulated case even when $d_{12}\sim 10^3\lambda_0$. The compensation for the weak coupling that decays with the geometric separation between the antennas, is introduced via a higher value of modulation index which is seen to be  increasing in this case.

\subsection{Power balance: discussion in light of the optimization non-convexity}
The optimization that is carried out to obtain Fig.~\ref{Fig2} is  non-convex and as such multiple optima points may be found, among which the globally best solution exists. Each of the local optima solutions exhibits different dynamics in terms of power balance between the RF port \#1, the LTV element port \#2, and the radiated power. To demonstrate that, in Fig.~\ref{Fig3}(a) we show the optimization results of the gain bandwidth on a subspace of parameters with fixed $d_{12}=0.5\lambda_0$. The optimal results are plotted as a function of $d_{13}/\lambda_0$. For the stationary, LTI, antenna array the optimal GBW is plotted in dashed blue line, while for the LTV antenna array with a continuous blue line. The required modulation for the optimal LTV setup is shown by the dotted reddish-brown line.  Clearly, in Fig.~\ref{Fig3}(a) two optimal points exist, first (global) at $d_{13}=0.11\lambda_0$, and second (local) at $d_{13}=0.255\lambda_0$. These two optimal points correspond to  distinct solutions of the optimization problem. The transition between these solutions is identified by a jump discontinuity in the require modulation, and in discontinuity in the derivative of the optimal GBW with respect to $d_{13}$. Three distinct solutions are shown in Fig.~\ref{Fig3}(a), denoted as Sol A, B, and C, and color-coded by blue, green, and red backgrounds, respectively.
The global and local optima points over the entire range of $d_{13}$ exist in the domains of Sol A and C, and exhibit different power balance dynamics. This is demonstrated in Fig.~\ref{Fig3}(b) and (c).
Fig.~\ref{Fig3}(b,c) depict, for the case $d_{12}=0.5\lambda_0$ and as a function of the detuning frequency $\delta\omega$, the radiated power $P_{\rm rad}$ at the first harmonic (continuous blue line), the total input power at source terminal \# 1, i.e., $P_1$ (dotted green line), the total input power pumped into the system at terminal \#2, i.e., $P_2$ (dash-dotted cyan), where the LTV inductor is attached,  and the total input power $P_{\rm in}=P_1+P_2$ as a function of the source frequency deviation $\delta \omega$ (dashed magenta line). Fig.~\ref{Fig3}(b) shows the power dispersion for  the global GBW optimum at $d_{13}=0.11 \lambda_0$, with $L_0/L_1=1.0788$, and $L_3/L_1=1.0146$,  whereas Fig.~\ref{Fig3}(c) shows the power dispersion for the local GBW optimum at $d_{13}=0.255 \lambda_0$, with $L_0/L_1=0.9297$, and $L_3/L_1=1.0018$. In all the examples in this work, $L_1=3.986\mu$H, is set to adjust the first, primary, dipole \#1 to resonate when isolated  at frequency $\omega_0$ corresponding to $\lambda_0=1$m.  By comparing $P_{\rm rad}$ and $P_{\rm in}$ it is easily noted that $(i)$ the radiation efficiency is approaching 100\%, $(ii)$ the pumping power injected by the modulation at port \# 2 is not evenly distributed with the frequency deviation $\delta_\omega$, and $(iii)$ for distinct optimum points with similar optimal GBW (as in the two points shown in Fig.~\ref{Fig3}(a)), the injected power by the modulation network is differently distributed as a result of the inherent wave feedback in the LTV system to enhance the gain-bandwidth over the stationary scheme.

%%%%%%%%%%%%%%
\begin{figure}[]
\centering
\includegraphics[width=0.45\textwidth]{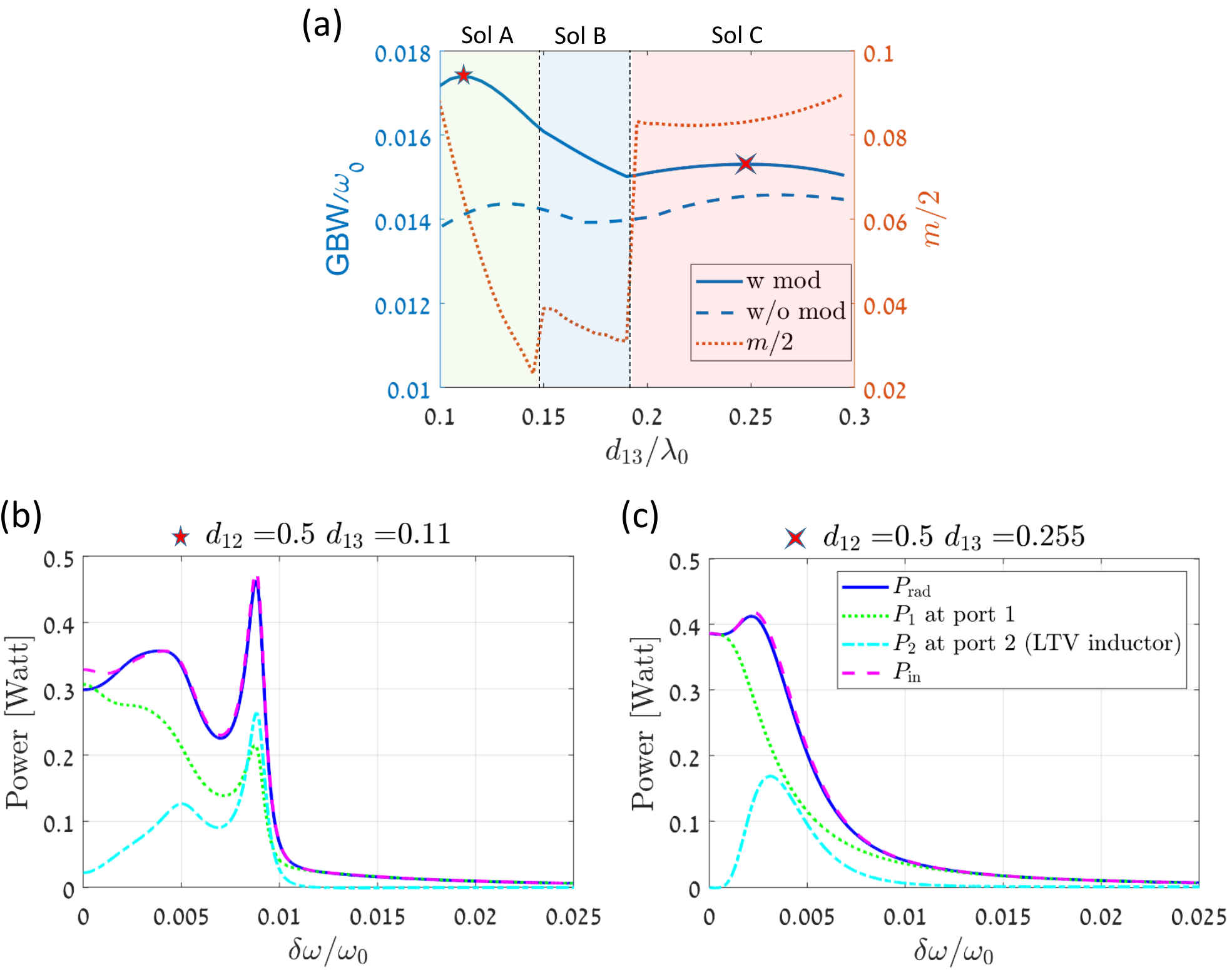}
\caption{(a) Optimal GBW over a subspace of parameters with $d_{12}=0.5\lambda_0$. Three distinct solutions are shown as a function of $d_{13}$. In these solutions, two optimal points exists, global, and local, exhibiting different power balance. (b) For the global optimum at $d_{13}=0.11 \lambda_0$, and (c) for the local optimum at $d_{13}=0.255 \lambda_0$. The radiated power $P_{\rm rad}$ in blue, the total input power at source terminal \# 1 ($P_1$ ) in green, the total input power pumped into the system at terminal \#2 ($P_2$) in cyan and the total input power $P_{in}=P_1+P_2$ in magenta as a function of the source frequency deviation $\delta \omega$.}
\label{Fig3}
\end{figure}
%%%%%%%%%%%%%%

The effect of the modulation of the inductor loading in antenna \#2 is achieved by pumping energy into (or out of) the antenna array and thus improving the GBW by increasing the gain and/or the bandwidth. In light of the non-convexity of the problem that gives rise to multiple possible solutions, it is of interest to explore the frequency dependence of the energy injection into the system at antenna \#2. Clearly, the bandwidth of the power injection by the LTV element is different than the overall bandwidth of the antenna system, and it is selected through the optimization in order to maximize the overall gain-bandwidth that is shown in Fig.~\ref{Fig2}. Fig.~\ref{Fig2b} depicts the effective mean square bandwidth of the injected power, $P_2$, (around the center frequency $\omega_0$) at the terminal of antenna \#2 at the optimal GBW as a function off the array size $d_{12}$.  While, for $d_{12} \lesssim 2$, the power pumping bandwidth varies significantly as $d_{12}$ changes in correlation with the modulation index, in Fig.~\ref{Fig2}, such to optimize the array response by widening the bandwidth and thus optimize GBW, for large arrays  with $d_{12}/\lambda_{12} \gtrsim 3$, the optimal solution dynamics varies, and the effective power pumping bandwidth decreases to zero, thus implying that the energy pumped into the system is concentrated in a small frequency range around $\omega_0$. Moreover, this decrease in the bandwidth to practically zero implies that the modulation has a negligible effect on the GBW and thus the GBW of the modulated array approaches eventually that of the non-modulated array (compare the solid and dotted blue lines in Fig.~\ref{Fig2}). This decrease in the pumping bandwidth is associated with an increase in the modulation index that further stems the ineffectiveness of the LTV modulation for extremely large arrays where the indirect coupling practically approaches zero. Nevertheless, for a very large range of array sizes, or equivalently put, for a very large range of indirect coupling cases, the use of the indirect modulation scheme may yield substantial enhancement in the overall performance, beyond what is expected by a stationary (LTI) antenna system.

\begin{figure}[]
\centering
\includegraphics[width=8cm]{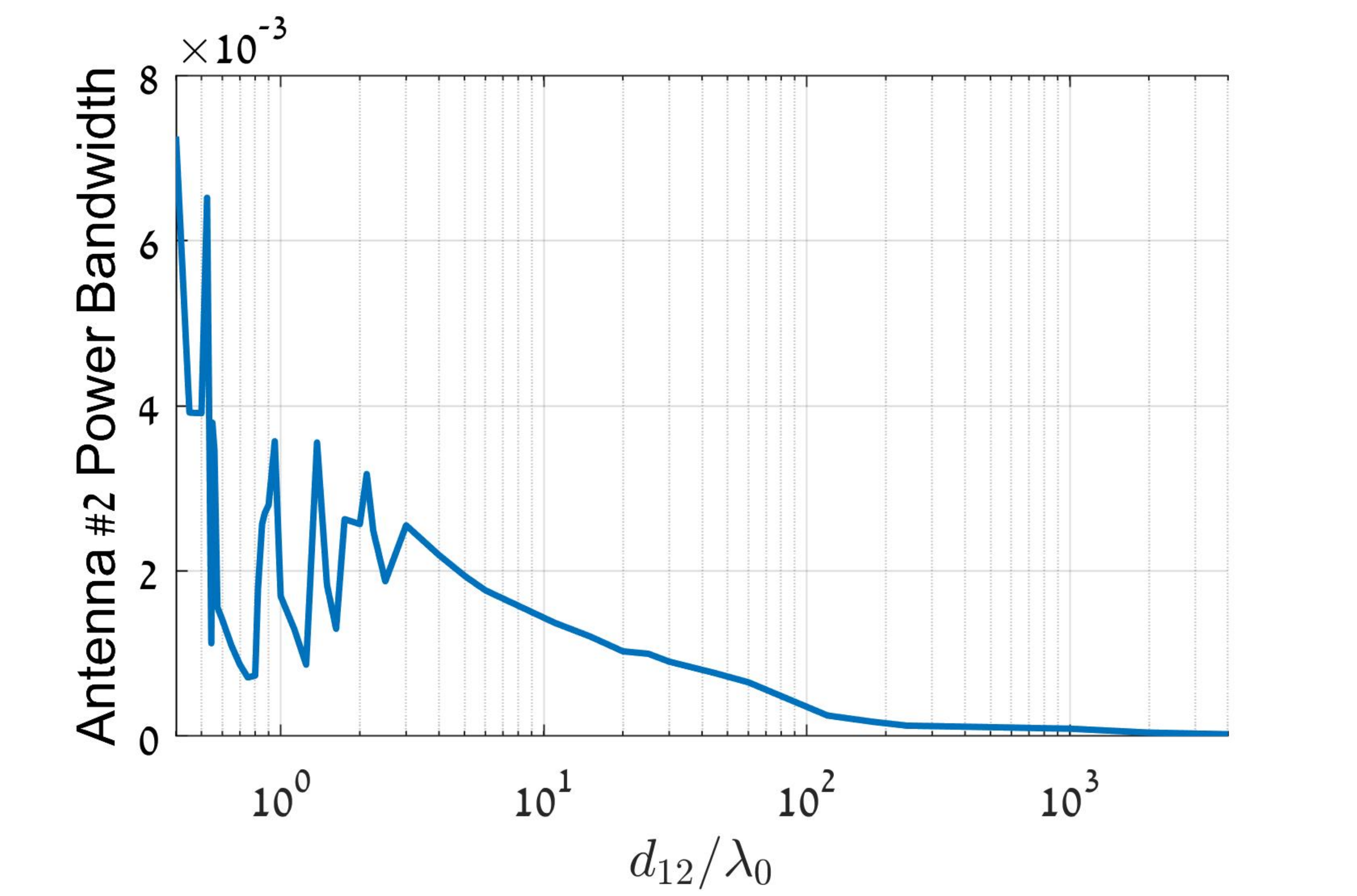}
\caption{The power bandwidth as seen from the time-modulated scatterer, i.e., antenna \#2.}
\label{Fig2b}
\end{figure}

%%%%%%%%%%%%%%%%%%%%%%%%
\section{Conclusions}
\label{Conclusions}
%%%%%%%%%%%%%%%%%%%%%%%%
In this paper we explored the possibility of indirect time modulation in antenna systems as a convenient way to go beyond the performance of stationary antenna systems without any actual modifications carried out on the original antenna configuration. The modulation is performed by the introduction of an additional time modulated parasitic element in a proximity to the stationary antenna, and thus it affects the overall radiation characteristics through the electromagnetic coupling. This effect is twofold, first on the feed-antenna matching, and second on the antenna radiation properties. This effect by an electromagnetic coupling which reduces with the distance is enhanced significantly through a feedback mechanism that essentially tunes to optimum the modulation depth per a given distance. Thus, interestingly, even at relatively large distances between the parasitic element and the other elements in the antenna system, the effect of the effective mutual coupling is prominent. We demonstrated the use of this effect for the enhancement of the gain-bandwidth of a 3-dipole Yagi-Uda-like antenna array. Although demonstrated on a particular antenna, our results are not limited for that configuration but may be applicable in other antenna systems as well.

%*****
\begin{appendices}
%*****

%%%%%%%%%%%%%%%%%%%%%%%%%%
\section{Physical model and mathematical formulation}
\label{Prelimenaries}
%\label{Layout}
%%%%%%%%%%%%%%%%%%%%%%%%%%
%The physical layout of the Yagi-Uda like antenna in the main text is that of an array of three identically short $\1z$ directed dipole antennas (indicated as $a_1
%$, $a_2$, $a_3$) with length $\ell_q$ that are aligned along the $\1x$ axis, see in Fig. 1(d) in the main text. The distance between $a_1$ and $a_2$ is $d_{12}$, between $a_1$ and $a_3$ is $d_{13}$, and between $a_2$ and $a_3$ is $d_{23}$ such that $d_{12}=d_{13}+d_{23}$ where in the present discussion, $d_{12} > d_{13}$.  Antenna $a_1$ is connected via a matching section composed of a single stationary inductor, $L_1$ to a transmission line (TL) with characteristic impedance $Z_0$, antenna $a_2$ is loaded with a time dependent (periodically modulated) inductor, $L_2(t)$ and antenna $a_3$ is loaded with stationary inductor $L_3$.
%
%Since the antennas are short (length $< \lambda/4$) their input impedance is essentially capacitive. The inductor $L_1$ is, thus, set to match the TL terminal when the system is composed of only antenna $a_1$ at some reference (central) frequency $\omega_r$, and $Z_0$ is set to equal the real part of the input impedance of antenna $a_1$. The inductance of $L_2(t)$, given in Eq.~(\ref{L2_ta}), is periodically modulated in time by
%\be
%L_2(t)=L_{20} \bigl (1+2\lambda_1 \cos(\omega_c t)\bigr)  = L_{20}\sum_{m=-1}^{1} \lambda_m e^{m\omega_c t}
%\label{L2_t}
%\ee
%with $|\lambda_{\pm1}|<0.5$, $\lambda_0=1$, and $\im\{\lambda_m\}=0$, and
%where $\omega_m=2\omega_r$ is the modulation frequency with $\omega_r$ a reference frequency.

We assume that the coupled three dipole system, including the matching inductance $L_1$, comprises a three-port network with a known admittance matrix $\bf{Y}(\omega)$. The ports' terminal points are shown by the orange dots in Fig.~\ref{Fig1}. We denote the terminal currents and voltages of the three ports by $\{V_1,V_2,V_3\}$ and $\{I_1,I_2,I_3\}$.  The currents are considered positive  when flowing into the terminals.
Note that this three-port network is LTI. The time invariance of the system is obtained only once the time-varying inductance $L_2(t)$ is connected at port \#2.
The voltage-current relation in the time-domain is obtained by the following convolution with the Fourier-transformed admittance ${\bf Y}(t)={\cal F}^{-1}\left\{{\bf Y}(\omega) \right\}$,
\be
\begin{bmatrix}
I_1(t)\\
I_2(t)\\
I_3(t)\\
\end{bmatrix}
=
\begin{bmatrix}
Y_{11}(t) & Y_{12}(t) & Y_{13}(t) \\
Y_{21}(t) & Y_{22}(t) & Y_{23}(t)\\
Y_{31}(t) & Y_{32}(t) & Y_{33}(t) \\
\end{bmatrix}
\label{Ymatrix_1}
\circledast
\begin{bmatrix}
V_1(t)\\
V_2(t)\\
V_3(t)\\
\end{bmatrix},
\ee
temporal convolution is indicated by $\circledast$ and the $\{Y_{m,n}(t)\}$, $m,n=1,2,3$ are the admittance matrix temporal operator elements of the system.

Eq.~(\ref{Ymatrix_1}) describes the relations between the voltages and currents dictated by the three-port network itself. However, the knowledge of the constrains obtained by the loading elements at the terminals of the three-port network are essential in order uniquely determine the system's state. We refer to these constraints as the constitutive relations by the loading elements.
Specifically, the constitutive relations between the voltages and currents across the time modulated $L_2$ and the stationary $L_3$ inductors at terminals $t_2$ and $t_3$, respectively, are given by:
\be
\label{t23_VI}
V_2(t)=-\frac{d}{dt}\left[ L_2(t) I_2(t)\right ], \qquad V_3(t)=-L_3\frac{dI_3(t)}{dt}
\ee
%where $\chi(t)$ is a convolution operator representing the current-voltage time domain transfer relation associate an LTI inductor, $\chi(t)=L_{30} \delta^{(1)}(t)$ where $\delta^{(1)}(t)$ indicates time domain differentiation \footnote{Note that a similar relation for the current-voltage time domain transfer relation associate an LTI capacitor is given by $\chi(t)=\frac{1}{C_{30}} \delta^{(-1)}(t)$, where $\delta^{(-1)}(t)$ indicates time domain differentiation.}
while on terminal $t_1$ of the system,
\be
\label{t1_VI}
V_1(t)=V_1'(t)+L_1\frac{d}{dt}I_1(t),
\ee
where $V_1'(t)$ is the voltage between antenna $a_1$ terminal points (note that terminals $t_2$ and $t_3$ are those of the antennas $a_2$ and $a_3$, respectively).

The feed port, \#1, is connected to a TL that provides the incoming signal wave $V_1^+$ which is assumed to be known. Thus, at this port, it is of interest to find the relation between the incoming wave $V_1^+$ and the reflected wave $V_1^-$, in addition to the relations between the currents and voltages at ports \#2 and \#3.
This can be achieved by using $V_1=V_1^++V_1^-$ and $I_1=(V_1^+-V_1^-)/Z_0$ in Eq.~(\ref{Ymatrix_1}), leading to

%
%
%
%Since at $t_1$ the transmission line is connected to the antenna system (actual source) is a transmission line terminal, there are also $V_1(t)=V_1^+(t)+V_1^-(t)$ and $I_1(t)=\left(V_1^+(t)-V_1^-(t)\right)/Z_0$ with $V_1^\pm(t)$ denore the incoming and reflected traveling voltage waves at the TL terminal side of $t_1$. Assuming the excitation is applied by the incoming $V_1^+(t)$ wave, \Eqref{Ymatrix_1} can be re-formulated formulated for $V_1^-(t)$, $I_2(t)$ and $I_3(t)$ as

\begin{equation}
\begin{aligned}
\begin{bmatrix}
\delta(t)+z_0Y_{11}(t) & 0 & 0 \\
-Y_{21}(t) & \delta(t) & 0\\
-Y_{31}(t) & 0 &\delta(t) \\
\end{bmatrix}
\circledast
\begin{bmatrix}
V_1^-(t)\\
I_2(t)\\
I_3(t)\\
\end{bmatrix}
&=
\\[1ex]
\begin{bmatrix}
\delta(t)-z_0Y_{11}(t) & -z_0 Y_{12}(t) & -z_0 Y_{13}(t) \\
Y_{21}(t) & Y_{22}(t) & Y_{23}(t)\\
Y_{31}(t) & Y_{32}(t) & Y_{33}(t) \\
\end{bmatrix}
\label{Ymatrix_2}
&\circledast
\begin{bmatrix}
V_1^+(t)\\
V_2(t)\\
V_3(t)\\
\end{bmatrix}
\end{aligned}
\end{equation}

where $\delta(t)$ is Dirac's delta which stands here as the identity temporal convolution operator.

%%%%%%%%%%%%%%%%%%%%%%%%%%
\section{Single tone (harmonic) excitations}\label{SingleTone}
%\label{HrmonicExcitation}
%%%%%%%%%%%%%%%%%%%%%%%%%%
Appendix \ref{Prelimenaries} above, provides the time domain formulation for the excitation problem for a general temporal form of an incoming impinging wave $V_1^+$. While it is general, it is implicit since the admittance operator ${\bf Y}(t)$ is in general an inconvenient function of time.
The analysis simplifies significantly once a single tone, time-harmonic, excitation is assumed. In this case, a quasi-frequency domain, harmonic-balance analysis can be performed.  This case is described below. Once this problem is solved, since the system is linear, the principle of superposition will be applied to solve a finite bandwidth excitation. See subsequent Appendix \ref{Twotone}.

Assuming the exciting incoming wave field is harmonic, $V_1^+(t)=V_0\cos(\omega t)$, and since the $L_2(t)$ inductance's modulation is also harmonic at $\omega_m$, the harmonic balance method can be used to transform the matrix convolution system in \Eqref{Ymatrix_2} into an algebraic system. To that end, first expanding:
\bse
\label{VI_expansion}
\bea
V_1^+(t)&=\half \sum_n V_{1_n}^+(\omega) e^{j\omega_n t} +c.c.,
\label{VI_expansion_a}
\\[1ex]
V_1^-(t)&=\half \sum_n V_{1_n}^-(\omega) e^{j\omega_n t} +c.c.
\label{VI_expansion_b}
\\[1ex]
I_2(t)&=\half \sum_n I_{2_n}(\omega) e^{j\omega_n t} +c.c.
\label{VI_expansion_c}
\\[1ex]
I_3(t)&=\half \sum_n I_{3_n}(\omega) e^{j\omega_n t} +c.c.
\label{VI_expansion_d}
\eea
\ese
where $c.c$ indicates complex conjugation, $V_{1_n}^+ =0$ for $|n|>0$,  $\omega_n=\omega+n\omega_m$,  and $V_{1_n}^-(\omega)$, $I_{2,3_n}(\omega)$ are sets of frequency dependent expansion coefficients. It should be indicated that since since all temporal quantities are real functions time, symmetry relations exists between the discrete frequency dependent expansion coefficients $V_{1_n}^-(\omega)$, $I_{2,3_n}(\omega)$.
					
Insertion of $I_2(t)$ of \Eqref{VI_expansion_b} into \Eqref{t23_VI} gives
\bse
\label{V23_dscrt}
\bea
V_2(t)&=-\frac{d}{dt} \left \{ \left [L_{0}\sum_{m=-1}^{1} \lambda_m e^{m\omega_m t}\right ] \right.
\nonumber
\\
&\qquad \qquad \qquad \times
\left. \left [ \half \sum_n I_{2_n}(\omega) e^{j\omega_n t} +c.c \right ]\right\}
\label{V23_dscrt_a}
\intertext{performing some algebraic manipulations, changing variables and rearranging gives rise to}
V_2(t)&= -\half \sum_n \sum_m j\omega_n L_{0}  \lambda_m I_{2_{n-m}}e^{j\omega_n t}+c.c.
\label{V23_dscrt_b}
\eea
Thus giving a representation for the $\{V_{2_n}\}$ set of coefficients in terms of the $\{I_{2_n}\}$ set of coefficients. Note that \Eqref{V23_dscrt_b} is a discrete convolution equation, and that  in light of Eq.~(\ref{L2_ta}), in Eq.~(\ref{V23_dscrt}) $\lambda_0=1$ whereas $\lambda_{\pm1}=m/2$.
\ese
Similarly, inserting $I_3(t)$ of \Eqref{VI_expansion_c} into the expression for $V_3(t)$ in \Eqref{t23_VI} and noting that convolution of a time dependent function $a(t)$ with the time harmonic exponential $e^{j\omega t}$ gives the $a(t) \circledast e^{j\omega t} = \8a(\omega) e^{j\omega t}$, where overtilde, $\8{ \ }$ frequency domain Fourier transform, eventually yields,
\begin{equation}
\begin{aligned}
\label{V3_dscrt}
V_3(t)&=-\half\sum_n I_{3_n}\8\chi(\omega_n)e^{j\omega_n t} +c.c.,
\\[1ex]
\8\chi(\omega_n)&= j\omega_n L_{3}
\end{aligned}
\end{equation}
Thus giving a representation for the set of coefficient $\{V_{3_n}\}$ in terms of the $\{I_{3_n}\}$ set of coefficients.

Writing explicitly each of the three equations in \Eqref{Ymatrix_2}, inserting the expressions of the voltages in \Eqref{VI_expansion} and using orthogonality of the harmonic exponential gives a transform of the convolution equation \Eqref{Ymatrix_2} into the algebraic equation:
\bse
\label{Ymatrix_3xx}
\be
\begin{bmatrix}
V_{1_n}^-\\
I_{2_n}\\
I_{3_n}\\
\end{bmatrix}
=
\begin{bmatrix}
M_{11}(\omega_n) & M_{12}(\omega_n) & M_{13}(\omega_n) \\
M_{21}(\omega_n) & M_{22}(\omega_n) & M_{23}(\omega_n)\\
M_{31}(\omega_n) & M_{32}(\omega_n) & M_{33}(\omega_n) \\
\end{bmatrix}
\label{Ymatrix_3a}
\begin{bmatrix}
V_{1_n}^+\\
V_{2_n}\\
V_{3_n}\\
\end{bmatrix}
\ee
where
\bea
&\begin{bmatrix}
M_{11}(\omega_n) & M_{12}(\omega_n) & M_{13}(\omega_n) \\
M_{21}(\omega_n) & M_{22}(\omega_n) & M_{23}(\omega_n)\\
M_{31}(\omega_n) & M_{32}(\omega_n) & M_{33}(\omega_n) \\
\end{bmatrix}
\nonumber
\\
&=
%\begin{bmatrix}
%1+z_0 \8y_{11}(\omega_n) & 0 & 0 \\
%-\8y_{21}(\omega_n) & 1 & 0\\
%-\8y_{31}(\omega_n) & 0 &1 \\
%\end{bmatrix} ^{-1}
\3A^{-1}
\begin{bmatrix}
1-z_0 \8y_{11}(\omega_n) & -z_0 \8y_{12}(\omega_n) & -z_0 \8y_{13}(\omega_n) \\
\8y_{21}(\omega_n) & \8y_{22}(\omega_n) & \8y_{23}(\omega_n)\\
\8y_{31}(\omega_n) & \8y_{32}(\omega_n) & \8y_{33}(\omega_n) \\
\end{bmatrix}
%\label{Ymatrix_3b}
\eea
where
\be
\3A = \begin{bmatrix}
1+z_0 \8y_{11}(\omega_n) & 0 & 0 \\
-\8y_{21}(\omega_n) & 1 & 0\\
-\8y_{31}(\omega_n) & 0 &1 \\
\end{bmatrix}.
\ee
\ese

Now that the convolution equation was transformed to its algebraic counterpart for the case of harmonic modulation and excitation, the expressions of the coefficients $\{V_{2,3_n}\}$ as follows from \Eqref{V23_dscrt} can be inserted into Eq.~(\ref{Ymatrix_3xx}) to give
\bse
\label{Ymatrix_3b}
\bea
&V_{1_n}^-+j\omega_n L_{2} M_{12}(\omega_n) \sum_m \lambda_m I_{2_{n-m}}
\nonumber
\\[1ex]
&\qquad \qquad \qquad+\8\chi(\omega_n) M_{13}(\omega_n) I_{3_n} =V_{1_0}^+ M_{11} \delta_{n0}
\label{Ymatrix_3b_1}
\\[1ex]
&I_{2_n}+j\omega_n L_{2} M_{22}(\omega_n) \sum_m \lambda_m I_{2_{n-m}}
\nonumber
\\[1ex]
&\qquad \qquad \qquad+ \8\chi(\omega_n) M_{23}(\omega_n) I_{3_n}=V_{1_0}^+ M_{21} \delta_{n0}
\label{Ymatrix_3b_2}
\\[1ex]
&j\omega_n L_{2} M_{32}(\omega_n) \sum_m \lambda_m I_{2_{n-m}}
\nonumber
\\[1ex]
&\qquad \qquad + \left [ 1 + \8\chi(\omega_n) M_{33}(\omega_n) \right ] I_{3_n}= V_{1_0}^+ M_{31} \delta_{n0}
\label{Ymatrix_3b_3}
\eea
where $\delta_{n0}$ is the Kronicker's delta, which equals $1$ only for $n=0$.
\ese

The set of coupled equations \Eqref{Ymatrix_3b} represents the dynamics of the modulated system. Its solution gives the the reflected voltage and currents expansion coefficients which depend on the physical layout and the modulation parameters (that are embedded in the $\lambda_m$, $\8 \chi$ and $M_{nm}$ parameters). Since the number of harmonics is infinite, it is impractical to solve \Eqref{Ymatrix_3b} for all of them therefore, it will be solved for a truncated set of harmonics with $0 \le |n| \le N_V$. Note that the $n=0$ harmonic is the dominant one. One approach to obtain the currents and voltage harmonic magnitude (``the solution'') is to recast \Eqref{Ymatrix_3xx} as a large block matrix. Alternative approach is to manipulate Eq.~(\ref{Ymatrix_3xx}) such to obtain a recursive equation for $\{I_{2_n}\}$ that once solved can be used to gives rise to $\{I_{3_n}\}$ and $\{V_{1_n}^-\}$. We will follow this second approach. To that end, multiplying \Eqref{Ymatrix_3b_2} by $M_{32}(\omega_n)$ and \Eqref{Ymatrix_3b_3} by $M_{22}(\omega_n)$ and subtracting yields:
\bse
\label{I_3_n_rec}
\be
\begin{aligned}
I_{3_n}&=\frac{M_{32}(\omega_n)}{M_{22}(\omega_n)-\8\chi(\omega_n) \mu_2(\omega_n)} I_{2_n}
\\[1ex]
&\qquad\qquad+\frac{\mu_1(\omega_n)}{\8\chi(\omega_n)\mu_2(\omega_n)-M_{22}(\omega_n)}\delta_{n0}V_{10}^+
\label{I_3_n_rec_1}
\end{aligned}
\ee
where
\bea
\mu_1(\omega_n)&=M_{32}(\omega_n)M_{21}(\omega_n)-M_{22}(\omega_n)M_{31}(\omega_n),
\label{I_3_n_rec_2}
\\[1ex]
\mu_2(\omega_n)&=M_{32}(\omega_n)M_{23}(\omega_n)-M_{22}(\omega_n)M_{33}(\omega_n).
\label{I_3_n_rec_3}
\eea
Note that due to the $\delta_{n0}$ at the rightmost term on the right hand side in \Eqref{I_3_n_rec_1}, this term is evaluated at $\omega_0=\omega$ the excitation frequency.
\ese
Inserting $I_{3_n}$ of  \Eqref{I_3_n_rec} into \Eqref{Ymatrix_3b_2} to obtain an equation for $I_{2_n}$
\bea
&\left [ 1+\8\chi(\omega_n) \frac{M_{32}(\omega_n)M_{23}(\omega_n)}{M_{22}(\omega_n)-\8\chi(\omega_n) \mu_2(\omega_n)}\right ]I_{2_n}
\nonumber
\\[1ex]
&\qquad\qquad\qquad+ j \omega_n L_{0} M_{22}(\omega_n) \sum_m \lambda_m I_{2_{n-m}}
\nonumber
\\[1ex]
 &=\left [ M_{21}(\omega_n)-\frac{\mu_1(\omega_n)}{\8\chi(\omega_n)\mu_2(\omega_n)-M_{22}(\omega_n)}\right ]\delta_{n0}V_{10}^+.
 \label{I_2_n_rec_1}
\eea
Recalling that the $m$ summation is extended over $m=\{-1,0,1\}$, see in \Eqref{L2_ta}, \Eqref{I_2_n_rec_1} can be rearranged as:
\be
\label{I_2_n_rec_3}
 A_n I_{2_{n-1}}+B_n I_{2_{n}}+C_n I_{2_{n+1}} = D_n\delta_{n0}V_{10}^+
\ee
where
\bse
\label{I_2_n_rec_2}
\bea
A_n&=j\omega_n L_{0} M_{22}(\omega_n) \lambda_1,
\label{I_2_n_rec_4}
\\[1ex]
B_n&=1+\8\chi(\omega_n) \frac{M_{32}(\omega_n)M_{23}(\omega_n)}{M_{22}(\omega_n)-\8\chi(\omega_n) \mu_2(\omega_n)}
\nonumber
\\[1ex]
&\qquad\qquad\qquad\qquad+ j\omega_n L_{0} M_{22}(\omega_n) \lambda_0,
\label{I_2_n_rec_5}
\\[1ex]
C_n&=j\omega_n L_{0} M_{22}(\omega_n) \lambda_{-1},
\label{I_2_n_rec_6}
\\[1ex]
D_n&=M_{21}(\omega_n)-\frac{\mu_1(\omega_n)}{\8\chi(\omega_n)\mu_2(\omega_n)-M_{22}(\omega_n)}.
\label{I_2_n_rec_7}
\eea
\ese
Alternatively, defining $N(\omega_n)=1 + \8\chi(\omega_n) M_{33}(\omega_n)$ in \Eqref{Ymatrix_3b_3}, multiplying \Eqref{Ymatrix_3b_2} by $N(\omega_n)$ and \Eqref{Ymatrix_3b_3} by $\8\chi(\omega_n) M_{23}(\omega_n)$ and subtracting gives a somewhat more favourable recursive equation for $\{I_{2_n}\}$ in the form of \Eqref{I_2_n_rec_3} but with
\bse
\label{I_2_n_rec_8}
\bea
A_n&=j\omega_n L_{0} \mu_5(\omega_n) \lambda_1,
\label{I_2_n_rec_9}
\\[1ex]
B_n&=N(\omega_n)+ j\omega_n L_{0} \mu_5(\omega_n) \lambda_0,
\label{I_2_n_rec_10}
\\[1ex]
C_n&=j\omega_n L_{0} \mu_5(\omega_n) \lambda_{-1},
\label{I_2_n_rec_11}
\\[1ex]
D_n&=\mu_6(\omega_n).
\label{I_2_n_rec_12}
\eea
where
\bea
&\mu_5(\omega_n)=M_{22}(\omega_n)N(\omega_n)-M_{32}(\omega_n)M_{23}(\omega_n) \8\chi(\omega_n),
\label{V_1_n_rec_2}
\\[1ex]
&\mu_6(\omega_n)=M_{21}(\omega_n)N(\omega_n)-M_{31}(\omega_n)M_{23}(\omega_n) \8\chi(\omega_n).
\eea
\ese
Finally, $\{I_{3_n}\}$, is obtained from \Eqref{Ymatrix_3b_2} to be given in Eq.~\eqref{I_3_n_rec_4}.
\begin{figure*}[!t]
\hrulefill
%\begin{strip}
%\hrulefill
\be
%\begin{align*}
%\bea
I_{3_n}=\frac{M_{21}(\omega_n)}{\8\chi(\omega_n) M_{23}(\omega_n)}V_{10}^+ \delta_{n0}
\label{I_3_n_rec_4}
- \frac{j\omega_n L_{0}M_{22}(\omega_n) \lambda_1 I_{2_{n-1}} + \left ( 1+j\omega_n L_{0}M_{22}(\omega_n) \lambda_0 I_{2_n}\right) + j\omega_n L_{0}M_{22}(\omega_n) \lambda_{-1} I_{2_{n+1}}}{\8\chi(\omega_n) M_{23}(\omega_n)}.
%I_{3_n}=&\frac{M_{21}(\omega_n)}{\8\chi(\omega_n) M_{23}(\omega_n)}V_{10}^+ \delta_{n0}
%\label{I_3_n_rec_4}
%\\[1ex]
%&- \frac{j\omega_n L_{20}M_{22}(\omega_n) \lambda_1 I_{2_{n-1}} + \left ( 1+j\omega_n L_{20}M_{22}(\omega_n) \lambda_0 I_{2_n}\right) + j\omega_n L_{20}M_{22}(\omega_n) \lambda_{-1} I_{2_{n+1}}}{\8\chi(\omega_n) M_{23}(\omega_n)}.
%\nonumber
%\eea
%\end{align*}
\ee
%\hrulefill
%\end{strip}
\hrulefill
\end{figure*}
Equation \eqref{I_2_n_rec_3} either  with \Eqref{I_2_n_rec_2} or \Eqref{I_2_n_rec_8}   is the key equation to be solved since by obtaining $I_{2_n}$, $I_{3_n}$ can be calculated using \Eqref{I_3_n_rec_1} and be inserted to \Eqref{Ymatrix_3b_1} to give $V_{1_n}^-$
\bse
\label{V_1_n_rec}
\be
\label{V_1_n_rec_1}
\begin{aligned}
V_{1_n}^-=\frac{M_{12}(\omega_n)}{M_{22}(\omega_n)} I_{2_n} - \8\chi(\omega_n)& \frac{\mu_3(\omega_n)}{M_{22}(\omega_n)} I_{3_n}
\\[1ex]
&+ \frac{\mu_4(\omega_n)}{M_{22}(\omega_n)} V_{1_0}^+ \delta_{n0},
\end{aligned}
\ee
\bea
\mu_3(\omega_n)&=M_{22}(\omega_n)M_{13}(\omega_n)-M_{12}(\omega_n)M_{23}(\omega_n),
\label{V_1_n_rec_2}
\\[1ex]
\mu_4(\omega_n)&=M_{22}(\omega_n)M_{11}(\omega_n)-M_{12}(\omega_n)M_{21}(\omega_n).
\eea
\ese

Equations \eqref{I_3_n_rec_1}, \eqref{I_2_n_rec_2} and \eqref{V_1_n_rec} can also be arranged differently in a manner that maybe more convenient for the calculation. Nevertheless, the structure of the recursion in \Eqref{I_2_n_rec_3} remains the same. Since $n$ is truncated $(-N_V \le n \le N_V)$, this recurision equation can be rearranged as a tridiagonal matrix equation:
\be
\3Z \, \3I_2 =\3V,
\label{Tri_Mat}
\ee
where $\3Z$ is an $(2N_V+1) \times (2N_V+1)$ tridiagonal matrix whose diagonal elements are $B_n$ for $n=-N_V, \ldots, 0, \ldots N_V$, the diagonal below the main diagonal is $A_n$ with with $n=-N_V+1, \ldots, 0, \ldots N_V$ and the diagonal above the main diagonal is $C_n$  with $n=-N_V, \ldots, 0, \ldots N_V-1$, $\3I_2=[ I_{2_{-N_V}}, I_{2_{-N_V+1}} , \dots I_{2_{N_V-1}}, I_{2_{-N_V}} ]^T$ where superscript  $T$ indicates transpose and $\3V$ is a column vector of zeros whose $N_V+1$ element (central element) is $D_0 V_{10}^+$. The matrix system in \Eqref{Tri_Mat} can be inverted to give the harmonic coefficient vector $\3I_2$ and from here, using \Eqref{I_3_n_rec_1} and \Eqref{V_1_n_rec}, the rest of the coefficients can be obtained. Once $\3I_2$ was calculated, the current at the terminals of antenna $a_3$, $\3I_3=[ I_{3_{-N_V}}, I_{3_{-N_V+1}} , \dots I_{3_{N_V-1}}, I_{3_{-N_V}}]^T$, and the voltage at the terminal of antenna $a_1$ and $\3V_1^-=[ V_{1_{-N_V}}^-, V_{1_{-N_V+1}}^- , \dots V_{1_{N_V-1}}^-, V_{1_{-N_V}}^-]^T$ can also be obtained via \Eqref{I_3_n_rec_4} and \Eqref{V_1_n_rec_1}.

%%%%%%%%%%%%%%%%%%%%%%%%%%
\section{Two-tone (finite-bandwidth) excitation}\label{Twotone}
\label{ModulatedExcitation}
%%%%%%%%%%%%%%%%%%%%%%%%%%
The calculation so far assumed a single tone harmonic excitation signal $V_1^+=V_0\cos(\omega t)$ in the vicinity of some center frequency $\omega_0$, i.e., $\omega=\omega_0+\delta_\omega$, $|\delta_\omega| \ll \omega_0$. A more practical case is of a modulated signal. In the following discussion we consider an AM symmetric two-tones signal of the form
\be
V_1^+=V_0 \cos(\omega_0 t ) \cos(\delta_\omega t)=\frac{V_0}{2} \bigl (\cos(\omega_a t)+\cos(\omega_b t) \bigl )
\label{AMsig_1}
\ee
with $\omega_a=\omega_0+\delta_\omega$ and $\omega_b=\omega_0-\delta_\omega $. See the single sided spectrum picture of $V_1^+$ in Fig.~\ref{Fig1}c above.
Obviously, $V_1^+$ here consists of a superposition of two single tone signals at frequencies $\omega_a$ and $\omega_b$.
Thus, we repeat the single tone solution of Appendix \ref{SingleTone} for these two frequencies, and find the corresponding $\3I_{2,3}$ and $\3V_1^-$. We shall denote the solution for each one of the tones, at $\omega_a$ and at $\omega_b$, with a subscript $a$ and $b$, respectively.

Assuming that $u(t)$ stands for either $V_1^-(t)$, $I_{2}(t)$, or $I_{3}(t)$, then the two tone solution reads $u(t)=u_a(t)+u_b(t)$,  where,
\be
\begin{aligned}
\label{ut_1}
u_a(t)=& \frac{1}{2} \sum_{n=-\infty}^{\infty} u_{a_n}e^{j\left( \omega_a+2n\omega_0 \right)t} +c.c.,
\\[1ex]
u_b(t)=& \frac{1}{2}\sum_{n=-\infty}^{\infty} u_{b_n}e^{j\left( \omega_b+2n\omega_0 \right)t}+c.c.
\end{aligned}
\ee
and with $u_{a_n}$ and $u_{b_n}$ denoting the Fourier coefficients of the signals. By some algebraic manipulations and using the symmetry of the excitation tone it follows that
\bea
\label{ut_1}
u(t)&= \frac{1}{2} \sum_{n=0}^\infty \left [u_{a_n}+u_{b_{-(n+1)}}^\ast \right] e^{j\left(\omega_a+2n\omega_0 \right)t} +c.c.
\nonumber
\\[1ex]
&+ \frac{1}{2} \sum_{n=0}^\infty \left [u_{b_n}+u_{a_{-(n+1)}}^\ast \right] e^{j\left(\omega_b+2n\omega_0 \right)t} +c.c.
\eea
with $*$ denoting a complex conjugate. Eq.~(\ref{ut_1}) gives the amplitude of the harmonic contributions at each of the frequencies and will be used below in the calculation of the power and energy of the signals.

\section{Derivation of the impedance matrix entries for three coupled antennas}
The self and mutual impedances of a pair of two parallel dipole antennas is given in the literature \cite{KrausAntennas}. We shall denote this 2 by 2 matrix with $\tilde{\bf Z}=[\tilde{Z}_{11},\tilde{Z}_{12};\tilde{Z}_{21}, \tilde{Z}_{22}]$.
In our setup, however, there are three adjacent dipole elements. Thus,  this system should be represented by a 3 by 3 impedance matrix that shall be denotes by ${\bf Z}$ (with no tilde above), with entries $Z_{ij}$ where $i,j=1..3$. Obviously, these entries are different from the impedance matrix entries of the two adjacent dipoles problem. Nevertheless, we show bellow how the two dipoles impedance matrix ${\tilde{\bf Z}}$ can be used in order to approximate, consistently, the three dipoles impedance matrix.

First, given the impedance matrix for the two-dipoles system, ${\tilde{\bf Z}}$, it can be associated with a corresponding scattering matrix ${\tilde{\bf S}}=[\tilde{S}_{11},\tilde{S}_{12};\tilde{S}_{21}, \tilde{S}_{22}]$, through the relation ${\tilde{\bf S}}=({\tilde{\bf Z}}-{\bf I})^{-1}({\tilde{\bf Z}}+{\bf I})$ \cite{Pozar}. So, with no loss of generality, we shall assume that the scattering matrix ${\tilde{\bf S}}$ is known.
Next, by using the flow chart in Fig.~\ref{ThreeAntennaSignalFlowChart}, we shall construct the entries $Z_{ij}$ for the three dipole system. Note that since here we derive the impedance matrix that describes the unloaded antenna configuration, that is, in the absence of the loading inductance $L_1,L_3$ and $L_2(t)$, the system is LTI and thus reciprocal (since all the materials are simple, with no gyrotropy, magnetization, etc.). This means that $Z_{12}=Z_{21}, Z_{13}=Z_{31}$ and $Z_{23}=Z_{32}$. In addition, while obviously in general $Z_{11}\neq Z_{22} \neq Z_{33}$, yet, there mathematical structure is very similar. Implying that once one of them is found, then, the remaining two can be immediately written by a careful change of indices. The same argument applies for the mutual impedance terms. Once $Z_{31}$ for example is found, $Z_{21}$  and $Z_{23}$ can be  immediately expressed.  In short, from the nine entries of the impedance matrix, we actually need to calculate only two. Below we derive in detail $Z_{31}$ and $Z_{11}$.

\begin{figure}[h]
    \begin{center}
       \includegraphics[width=80mm]{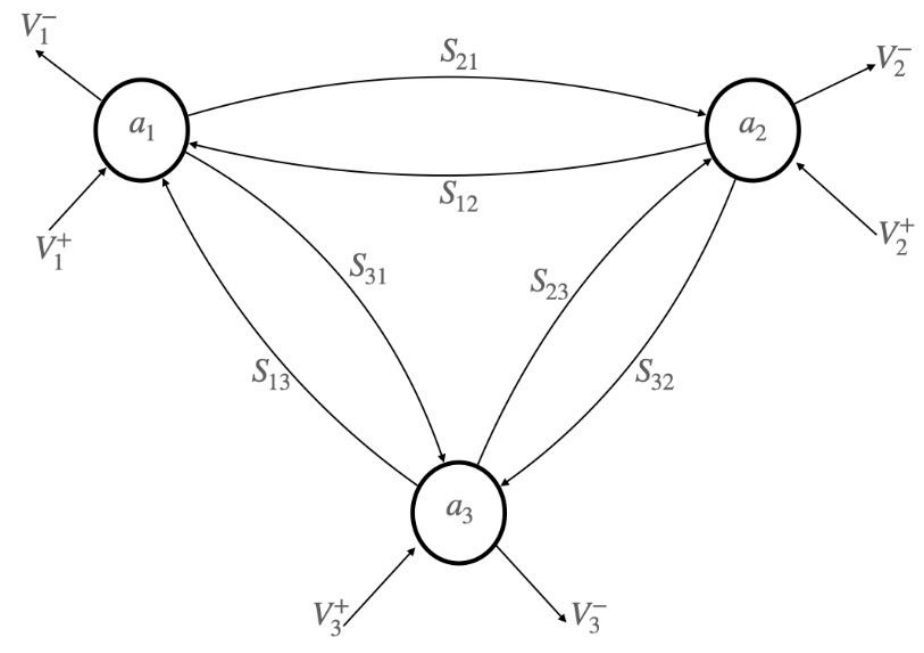}
    \end{center}
    \caption{Signal flow describing the mutual coupling interaction between the three antennas. The scattering parameters $S_{ij}$ are the scattering matrix terms of a pair of short dipole antennas. These coefficients are analytically known \cite{} or can be calculated numerically for more complicated antenna configurations.}
    \label{ThreeAntennaSignalFlowChart}
\end{figure}

\emph{Derivation of $Z_{31}$}. By definition,
\begin{equation}
Z_{31}=\left.\frac{V_3}{I_1}\right|_{I_2=I_3=0}
\end{equation}
The condition $I_2=I_3=0$, implies open boundary condition in ports \#2 and \#3. In terms of forward and backward waves, in ports \#1 and \#3, we have
\begin{equation}\label{Z31Def}
Z_{31}=\frac{V_3^++V_3^-}{(V_1^+-V_1^-)/Z_0}=\frac{2V_3^-}{(V_1^+-V_1^-)/Z_0}
\end{equation}
The second equality originates from the fact that port \#3 is open and therefore when looking from the port, outward, the voltage reflection coefficient is 1, i.e., $V_3^+=1\cdot V_3^-$.

Now, we shall express $V_3^-$ in terms of the incoming voltage wave at port \#1, that is $V_1^+$, and using the scattering matrix coefficients between the three \emph{isolated dipole pairs}, $1\leftrightarrow 2$, $2\leftrightarrow 3$, $2\leftrightarrow 3$.

\begin{eqnarray}\label{EqV3minus_derivation}
% \nonumber % Remove numbering (before each equation)
  V_3^- &=&   \tilde{S}_{31}V_1^++\tilde{S}_{21}\cdot1\cdot \tilde{S}_{32}V_1^+ \nonumber \\
  &&V_3^+\left[\tilde{S}_{23}\cdot1\cdot \tilde{S}_{32} + \tilde{S}_{31}\Gamma_1 \tilde{S}_{13} + \tilde{S}_{33} +    \right.\nonumber \\
  && \left. \tilde{S}_{23}\cdot1\cdot \tilde{S}_{12}\cdot\Gamma_1\cdot \tilde{S}_{31} + \tilde{S}_{13}\cdot\Gamma_1\cdot \tilde{S}_{21}\cdot 1 \cdot \tilde{S}_{32} \right]\nonumber\\
\end{eqnarray}

Let us explain the construction of Eq.~(\ref{EqV3minus_derivation}) using the signal flow diagram in Fig.~\ref{ThreeAntennaSignalFlowChart}. The outward voltage wave at port \#3, $V_3^-$, consists of a superposition of different contributions. First, it consists of the inward voltage wave at port \#1 is contributes to $V_3^-$ by the scattering coefficient $\tilde{S}_{31}$ when  the two  dipoles, 1 and 3, are isolated. Second, the inward voltage $V_1^+$ excites a outward voltage $V_2^+$ at port \#2 with coupling coefficient $\tilde{S}_{21}$ of the isolated pair of dipoles, 1 and 2. Then, it is fully reflected into the port (reflection coefficient $+1$), since the assumed boundary condition at port \#2 is open (recall, we set $I_2=0$ for this calculation). The inward wave, $V_2^+$ now contributes additionally to port \#3 via $\tilde{S}_{32}$. This leads to the second term in the expression.
Furthermore, we note that port \#3 is open, therefore, the outward wave $V_3^-$ is fully reflected at the port into an inward wave $V_3^+$, such that $V_3^+=V_3^-$ at the port. Consequently, additional contribution paths should be included for the sake of self consistency.  For instance, $V_3^+$ couples via $\tilde{S}_{23}$ to the outward wave at port \#2, $V_2^+$, which  is fully reflected with reflection coefficient $+1$ into an inward wave at port \#2 (which is open). The latter is back scattered via $\tilde{S}_{32}$ into the outward wave in port \#3, i.e., $V_3^-$. This leads to the first term in the rectangular brackets. Next, we have the inward wave $V_3^+$ that couples to the outward wave $V_1^-$ via $\tilde{S}_{13}$, then it is reflected back into an inward wave $V_1^+$. It is important to note that as opposed to ports \#2 and \#3 that are assumed open, the current in port \#1 is not zero. Therefore, the reflection coefficient of $V_1^-$ into $V_1^+$ should be carefully calculated,
\begin{equation}
\Gamma_1=\frac{V_1^+}{V_1^-}=\frac{Z_0-\tilde{Z}_{11}'}{Z_0+\tilde{Z}_{11}'}
\end{equation}
Then, the inward wave $V_1^+$ couples back to $V_3^-$ in port \#3 by $\tilde{S}_{31}$. This leads to the second term in the rectangular brackets.
The process continues, with additional multiple scattering paths, until self-consistency is achieved. The final result is given in the expression, the remaining derivation follows what was already outlined and for the sake of brevity it is left for the reader.

Using Eq.~(\ref{EqV3minus_derivation}) and in light of the discussion below it, simple rearrangement yields the connection between $V_3^-$ and $V_1^+$ using the isolated dipole pair scattering coefficients,
\begin{equation}\label{V3minus}
V_3^-=\frac{\tilde{S}_{13}+\tilde{S}_{12}\tilde{S}_{23}}{1-\tilde{S}_{23}^2-\Gamma_1(\tilde{S}_{13}^2+2\tilde{S}_{23}\tilde{S}_{12}\tilde{S}_{13})-\tilde{S}_{33}} V_1^+
\end{equation}
Following a similar signal flow approach as carried for the derivation of Eq.~(\ref{EqV3minus_derivation}), we can also express $V_1^-$ in terms of $V_1^+$, given that ports \#2 and \#3 are open. This reads,

\begin{equation}\label{V1minus}
V_1^- = V_1^+ \left[\tilde{S}_{11} + \tilde{S}_{21}^2 + \tilde{S}_{31}^2 + 2\tilde{S}_{21}\tilde{S}_{32}\tilde{S}_{13}\right]
\end{equation}
By using the expressions for $V_1^-$ and $V_3^-$ in Eq.~(\ref{V3minus}) and Eq.~(\ref{V1minus}) in the impedance $Z_{31}$ definition in Eq.~(\ref{Z31Def}), we obtain, Eq.~(\ref{Z_31})
%\begin{equation}
%Z_{31}=2Z_0 \frac{\left[\frac{\tilde{S}_{13}+\tilde{S}_{12}\tilde{S}_{23}}{1-(\tilde{S}_{33}+\tilde{S}_{23}^2+\Gamma_1(\tilde{S}_{13}^2+2\tilde{S}_{23}\tilde{S}_{12}\tilde{S}_{13}))}\right]}{1-[\tilde{S}_{11}+\tilde{S}_{21}^2+\tilde{S}_{31}^2+2\tilde{S}_{21}\tilde{S}_{32}\tilde{S}_{13}]}
%\end{equation}

\emph{Derivation of $Z_{11}$}.  The derivation of $Z_{11}$ follows very similar analytical streamline.
First, we calculate the impedance in the absence of the inductance $L_1$. We shall denote it by $Z_{11}'$. By definition,
\begin{equation}\label{Z11Def}
Z_{11}'=\left.\frac{V_1}{I_1}\right|_{I_2=I_3=0}
\end{equation}
We use Eq.~(\ref{V1minus}) to express $V_1=V_1^++V_1^-$, and $I_1=(V_1^+-V_1^-)/Z_0$, in terms of $V_1^+$ only. Then, plugging into Eq.~(\ref{Z11Def}) leads immediately to $Z_{11}'$ in Eq.~(\ref{Z_11}).
Next, because of the presence of the serial inductor, the impedance $Z_{11}$ seen from the TL into the  antenna including the inductor is given by $Z_{11}$ in Eq.~(\ref{Z_11}).

\emph{Derivation of the remaining impedance matrix entries}.
As discussed above, in light of the mathematical symmetry of the problem, the remaining impedance matrix entries in Eq.~(\ref{ImedanceMatrix}) are found by simple proper indices switching.

Finally, the  entries of the 3 by 3 impedance matrix are given below.

\bse
\bea
Z_{11}&=Z_{11}'+j\omega L_1\nonumber\\
Z_{11}'&=Z_0 \frac{1 +\left[\tilde{S}_{11}+\tilde{S}_{31}^2+\tilde{S}_{21}^2 +2\tilde{S}_{21} \tilde{S}_{32} \tilde{S}_{13}\right]}{1- \left[ \tilde{S}_{11}+\tilde{S}_{21}^2+\tilde{S}_{31}^2+2\tilde{S}_{21} \tilde{S}_{32} \tilde{S}_{13} \right]}
\label{Z_11}
\\[1ex]
Z_{21}&=2Z_0 \frac{ \left [ \frac{\tilde{S}_{21}+\tilde{S}_{31}\tilde{S}_{23}}{1-\left(\tilde{S}_{22}+\tilde{S}_{23}^2+\Gamma_1\left(\tilde{S}_{21}^2+2\tilde{S}_{23}\tilde{S}_{12}\tilde{S}_{31}\right) \right)}\right]}{1- \left[ \tilde{S}_{11}+\tilde{S}_{21}^2+\tilde{S}_{31}^2 +2\tilde{S}_{21} \tilde{S}_{32} \tilde{S}_{13} \right]},
\label{Z_21}
\\[1ex]
Z_{31}&=2Z_0 \frac{ \left [ \frac{\tilde{S}_{13}+\tilde{S}_{12}\tilde{S}_{23}}{1-\left(\tilde{S}_{33}+\tilde{S}_{23}^2+\Gamma_1\left(\tilde{S}_{13}^2+2\tilde{S}_{23}\tilde{S}_{12}\tilde{S}_{13}\right) \right)}\right]}{1- \left[ \tilde{S}_{11}+\tilde{S}_{21}^2+\tilde{S}_{31}^2 +2\tilde{S}_{21} \tilde{S}_{32} \tilde{S}_{13} \right]}, %\qqiad \qquad \Gamma_1=\frac{Z_0-Z_{11}}{Z_0+Z_{11}}
\label{Z_31}
\\[1ex]
Z_{22}&=Z_0 \frac{1 +\left[\tilde{S}_{22}+\tilde{S}_{12}^2+\tilde{S}_{23}^2 +2\tilde{S}_{21} \tilde{S}_{31} \tilde{S}_{32}\right]}{1- \left[ \tilde{S}_{22}+\tilde{S}_{12}^2+\tilde{S}_{23}^2+2\tilde{S}_{21} \tilde{S}_{31} \tilde{S}_{32} \right]}
\label{Z_22}
\\[1ex]
Z_{32}&=2Z_0 \frac{ \left [ \frac{\tilde{S}_{23}+\tilde{S}_{31}\tilde{S}_{12}}{1-\left(\tilde{S}_{33}+\tilde{S}_{31}^2+\Gamma_2\left(\tilde{S}_{32}^2+2\tilde{S}_{32}\tilde{S}_{21}\tilde{S}_{13}\right) \right)}\right]}{1- \left[ \tilde{S}_{22}+\tilde{S}_{12}^2+\tilde{S}_{23}^2 +2\tilde{S}_{21} \tilde{S}_{31} \tilde{S}_{32} \right]},
\label{Z_32}
\\[1ex]
Z_{33}&=Z_0 \frac{1 +\left[\tilde{S}_{33}+\tilde{S}_{31}^2+\tilde{S}_{32}^2 +2\tilde{S}_{31} \tilde{S}_{12} \tilde{S}_{21}\right]}{1- \left[ \tilde{S}_{33}+\tilde{S}_{31}^2+\tilde{S}_{32}^2+2\tilde{S}_{31} \tilde{S}_{12} \tilde{S}_{23} \right]}
\label{Z_33}
\eea
\label{ImedanceMatrix}
\ese

Note that in the $Z_{11}$ entry we have included the impedance of the serial inductor $L_1$ that is used for the matching of dipole \#1 when isolated. Thus, for this dipole, the terminal to which we refer is the point where the feed TL is connected. Furthermore, the reflection coefficient $\Gamma_2$ reads,
\begin{equation}
\Gamma_2=\frac{Z_0-Z_{22}}{Z_0+Z_{22}}.
\end{equation}
Lastly, note that in the derivations in Appendix A above we used the admittance rather than the impedance matrix of the three-dipole antenna system. They are trivially related by
\begin{equation}
{\bf Y}(\omega) = {\bf Z}^{-1}(\omega).
\end{equation}

%*****
\end{appendices}
%*****

% use section* for acknowledgment
\section*{Acknowledgment}
The authors would like to thank Dr. Arthur Yaghjian, and Prof. Andrea Alu for useful discussions.

\end{document}